\documentclass[conference]{IEEEtran}
\makeatletter
\def\ps@headings{%
\def\@oddhead{\mbox{}\scriptsize\rightmark \hfil \thepage}%
\def\@evenhead{\scriptsize\thepage \hfil \leftmark\mbox{}}%
\def\@oddfoot{}%
\def\@evenfoot{}}
\makeatother
\usepackage{graphicx}
\usepackage{cite}
\usepackage{latexsym}
\usepackage{pifont}

\usepackage{textcomp}
\usepackage{marvosym}
\usepackage{amssymb}
\usepackage{subfigure}
\ifCLASSINFOpdf
\else
\fi
\usepackage[cmex10]{amsmath}

\begin{document}
%
\title{QoS-Aware Joint Policies in Cognitive Radio Networks}
%
%
%
\author{\IEEEauthorblockN{Saber Salehkaleybar, Seyyed Arash Majd and Mohammad Reza Pakravan}
\IEEEauthorblockA{School of Electrical Engineering, Sharif University of Technology, Tehran, Iran\\
E-mails: saber\textunderscore saleh@ee.sharif.edu, arash\textunderscore majd@ee.sharif.edu, pakravan@sharif.edu}}

\maketitle

\begin{abstract}
One of the most challenging problems in Opportunistic Spectrum Access (OSA) is to design channel sensing-based protocol in multi secondary users (SUs) network. Quality of Service (QoS) requirements for SUs have significant implications on this protocol design. In this paper, we propose a new method to find joint policies for SUs which not only guarantees QoS requirements but also maximizes network throughput. We use Decentralized Partially Observable Markov Decision Process (Dec-POMDP) to formulate interactions between SUs. Meanwhile, a tractable approach for Dec-POMDP is utilized to extract sub-optimum joint policies for large horizons. Among these policies, the joint policy which guarantees QoS requirements is selected as the joint sensing strategy for SUs. To show the efficiency of the proposed method, we consider two SUs trying to access two-channel primary users (PUs) network modeled by discrete Markov chains. Simulations demonstrate three interesting findings: 1- Optimum joint policies for large horizons can be obtained using the proposed method. 2- There exists a joint policy for the assumed QoS constraints. 3- Our method outperforms other related works in terms of network throughput.
\end{abstract}
  


%
\section{Introduction}
\PARstart{W}{ith} the advent of the new applications in wireless data networks, bandwidth demand has increased, intensively. The majority of the usable frequency spectrum for wireless networks has already been assigned to licensed users. In contrast to the apparent spectrum scarcity, extensive measurements indicate that a large portion of licensed spectrum lies unused \cite{McHenry}. Thus, there is an intensive research attempt to present new techniques to utilize the unoccupied resources, efficiently \cite{Pollin,Prasad,Sadler}. To get higher frequency reuse efficiency, SUs should dynamically access PUs' channels. This concept is known as Opportunistic Spectrum Access (OSA) in literature\cite{Zhao}. In cognitive radio networks, channel occupation can be caused by two effects\cite{Fu}: One is the disturbance due to PUs' activities which can be modeled by finite state Markov chain\cite{Zhang}. The other is the impact caused by other SUs' transmission\cite{Poor}. 

Zhao et al. considered the Partially Observable Markov Decision Process (POMDP) framework for spectrum access \cite{Zhao}. They used POMDP approach to find an optimum policy for a single SU case. To generalize this solution to multi SUs, the simple {\em Carrier Sense Multiple Access/Collision Avoidance} (CSMA/CA) protocol was employed\cite{Zhao}. In another related work \cite{Liu1}, it is assumed that SUs obtain similar observations of PUs' channels and therefore converge to the same opportunity assessment if they employ single user strategy. Results demonstrate that applying optimal single user strategy to multi-user setting causes significant degrade in network throughput performance \cite{Liu1}. To overcome this problem, it was shown that, using a randomized policy selection, the network performance can be improved. However, this policy does not guarantee QoS requirements among competing SUs. Besides, SUs' collisions history is not considered in deriving belief vector. In other related work, Liu et al. considered two interfering SUs in a two-channel primary network\cite{Liu2}. Each SU observes different spectrum opportunity on each channel because of assumed structure of PUs' network. They proposed a myopic policy for SUs in which both SUs exchange their beliefs in each time slot. Simulations illustrate that this myopic policy achieves near-optimal network throughput.

Considering time-invariant spectrum opportunities, several works have been done using game theory\cite{Ji,Suris}. Recently, Fu and van der Schaar utilized stochastic games to present a solution for dynamic interaction among competing SUs \cite{Fu}. A Central Spectrum Moderator (CSM) is required in this model whose task is to announce the state of all channels to SUs in each time slot. However, having a centralized moderator is not practical in some cases and SUs can not sense all of channels in a limited time of single slot. In \cite{Pham}, Pham et al. proposed a game theoretic approach to QoS-aware channel selection for SUs which maximizes network throughput. They assumed that SUs know the spectrum availability before selecting appropriate channel.
 
Partially Observable Stochastic Game (POSG) is a general framework to solve multi-agent decision process \cite{Hansen}. In POSG, the state of the channel changes based on a discrete Markov Model and is partially observable to all agents. In this framework, each agent tries to maximize its own reward function in a repeated game. Hansen et al. proposed a Dynamic Programming (DP) approach to solve the problem of POSG. As a special case of POSG, the Dec-POMDP framework was investigated in \cite{Hansen} and \cite{Szer}, using the DP algorithm. In Dec-POMDP, all agents try to maximize a common reward function. Solving a Dec-POMDP problem by the DP algorithm becomes intractable when the horizon length of decision process increases.  For instance, the DP algorithm runs out of memory even for a small horizon length in a trivial example \cite{Saleh}. Seuken and Zilberstein developed Memory Bounded Dynamic Programming (MBDP) to overcome the time complexity of existing DP algorithm\cite{seuken}. 

\begin{figure}[!t]
\centering
\includegraphics[width=3.5in]{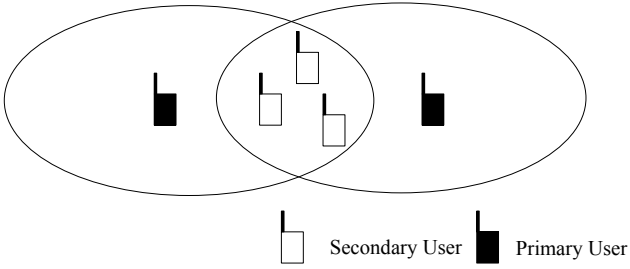}
\caption{Cognitive Radio Network}
\label{fign1}
\end{figure}

In this paper, our goal is to design QoS-aware joint policies for sensing decisions of SUs in order to maximize network throughput. It is assumed that PUs' network is slotted and all SUs have same spectrum opportunities in each time slot (see Fig. \ref{fign1}, SUs are in the transmission range of all PUs and different SUs have same observations of a specific channel). At first sight, to maximize the network throughput, each SU should be assigned a channel to exploit spectrum opportunities. Therefore, SUs avoid collisions with each other. This scheme is called partitioning strategy \cite{Liu2}. However, this strategy does not guarantee QoS requirements for SUs. For instance, when the probability of idle state is not the same for all channels, this strategy does not satisfy fairness. We formulate multiple SUs' joint policies by Dec-POMDP. In the proposed method, the MBDP algorithm is employed to find optimum or sub-optimum joint policies for large horizon length. A joint policy which guarantees QoS requirement is selected to obtain sensing strategy for SUs. To the best of our knowledge, the problem of synchronizing SUs for multi user setting, in the presence of collisions, have received little attention. The proposed method ensures transceiver SUs synchronization. To demonstrate the efficiency of the proposed method, we consider two SUs trying to access two-channel PUs' network modeled by discrete Markov chains. It is assumed that SUs have perfect sensing capability. Simulations yield these interesting findings: The MBDP algorithm obtains optimum solution for mentioned scenario. It is interesting to note that this algorithm is an approximate solution for Dec-POMDP and it does not guarantee to find an optimum joint policy. Moreover, there exists a joint policy which satisfies QoS constraints considered in simulations. Finally, comparing with two other related works \cite{Liu1,Liu2}, we find out that the proposed method outperforms \cite{Liu1,Liu2} in terms of network throughput.      

This paper is organized as follows. In Section II, we give an overview on Dec-POMDP. We also review the MBDP algorithm for Dec-POMDP. In Section III, the system model and Dec-POMDP formulations for cognitive radio network are described. In Section IV, we propose our method to extract QoS-aware joint policies. In Section V, as an example of our Dec-POMDP formulation, we define a scenario with two SUs trying to access a two-channel PUs' network. Also, the numerical simulation and results are provided. Finally, the conclusion is presented.     
\section{Definitions and Preliminaries}
In this section, we briefly review finite-horizon Dec-POMDP framework and the MBDP solution proposed for handling intractability problem of the DP algorithm. More details on the DP and MBDP algorithms could be found in \cite{Hansen,Szer,seuken}.
\subsection{Decentralized Partially Observable Markov Decision Process}
A Dec-POMDP is a tuple $\langle I,S,{b^0},{A_i},{O_i},P,{R}\rangle$, where,
\\-$I$ is a finite set of SUs indexed 1,...,n.
\\-$S$ is a finite set of states.
\\-$b^0\in \Delta(S)$ represents the initial state distribution.
\\-$A_i$ is a finite set of actions available to SU $i$ and $\vec{A}=$\texttimes$_{i\in I}A_i$ is the set of joint actions, where $\vec{a}=\langle a_1,...,a_n\rangle$ denotes a joint action.
\\-$O_i$ is a finite set of observations for SU $i$ and $\vec{O}=$\texttimes$_{i\in I}O_i$ is the set of joint observations, where $\vec{o}=\langle o_1,...,o_n\rangle$ denotes a joint observation.
\\-$P$ is the set of Markovian state transition and observation probabilities, where $P(s^{\prime},\vec{o}|s,\vec{a})$ denotes the probability that choosing joint action $\vec{a}$ in state $s$ yields a transition to the state $s^{\prime}$ and the joint observation $\vec{o}$.
\\-$R:S$\texttimes$\vec{A}$\ding{213}$\mathbb{R}$ is a reward function which depends on joint actions and current state.

Dec-POMDP may be defined over a finite or infinite sequence of stages. In this paper, we focus on finite horizon Dec-POMDP. At each stage, all SUs simultaneously select an action and receive an observation. The reward for SUs is computed based on their action and state of channels. The goal is to maximize the expected sum of rewards:
\begin{equation}
R^T=\mathbb{E}\left\{\displaystyle\sum_{t=1}^TR(t)\right\}
\end{equation} 
\subsection{Memory Bounded Dynamic Programming for Dec-POMDP}
Solving a Dec-POMDP means finding a joint policy that
maximizes the expected total reward. A policy for a single
agent $i$ can be represented by a decision tree $q_i$, where
nodes are labeled with actions and arcs are labeled with
observations (a so called a policy tree). If $Q^t_i$ denotes a
set of horizon-$t$ policy trees for agent $i$, a solution to Dec-POMDP with horizon $t$ can then be seen as a vector
of horizon-$t$ policy trees $\delta^t=(q_1^t,...,q_n^t)$ (a so called joint policy tree) where $q_i\in Q_i^t$. These policy trees can be constructed in two different approaches: top-down or bottom-up.

The first algorithm for solving Dec-POMDPs used a bottom-up approach \cite{Hansen}. Policy trees are constructed incrementally which means that the algorithm starts at the frontiers and works its way up to the roots using the DP algorithm. The DP algorithm updates in two steps. In the first step, the DP operator is given by a set $Q^t$ of depth-t policy trees. A set of depth-$t$ + 1 policy trees, $Q^{t+1}$, is generated by considering any depth-$t$ + 1 policy tree that makes
a transition after an action and observation to the root node of
depth-t policy tree in $Q^t$. This step is called exhaustive backup
\cite{Hansen}. In exhaustive backup, $|A||Q^t|^{|O|}$ depth-$t$ + 1 policy trees
are created. It is clear that the total number of constructed trees in each step increases exponentially. To alleviate this problem, unnecessary trees are pruned \cite{Hansen}. However, this modified DP algorithm runs out of memory even for simple problems \cite{seuken}.
\begin{table}[!t]
\renewcommand{\arraystretch}{1.3}
\caption{The pseudocode for the MBDP Algorithm\cite{seuken}}
\label{table1}
\centering
\begin{tabular}{|l|}
\hline
\\{\bf The MBDP Algorithm}
\\
\hline
\\{\bf begin}
\\1 \hspace{0.25cm} $maxTrees$ $\longleftarrow$ max number of trees before backup
\\2 \hspace{0.25cm} $T$ $\longleftarrow$ horizon of the Dec-POMDP
\\3 \hspace{0.25cm} $B$ $\longleftarrow$ pre-compute relevant belief for each horizon
\\4 \hspace{0.25cm} $Q_i^1, Q_j^1$ $\longleftarrow$ initialize 1-step policy trees for each SU 
\\5 \hspace{0.25cm} {\bf for} $t$ = $1$ $to$ $T$ {\bf do}
\\6 \hspace{0.75cm} $Q_i^{t+1}, Q_j^{t+1}$ $\longleftarrow$ Backup($Q^t_i$), Backup($Q^t_j$)
\\7 \hspace{0.75cm} $Sel_i^{t+1}, Sel_j^{t+1}$ $\longleftarrow$ empty
\\8 \hspace{0.75cm} {\bf for} $k$ = $1$ $to$ $maxTrees$ {\bf do}
\\9  \hspace{1.25cm} choose relevant belief ($b\in B$) for horizon $t$
\\10 \hspace{1.25cm} {\bf for each} $q_i\in Q_i^{t+1}$, $q_j \in Q_j^{t+1}$ {\bf do}
\\11 \hspace{1.75cm} evaluate each pair ($q_i$, $q_j$) with respect to b
\\12 \hspace{1.25cm} {\bf end}
\\13 \hspace{1.25cm} add best policy trees to $Sel^{t+1}_i$ and $Sel^{t+1}_j$
\\14 \hspace{1.25cm} delete these policy trees from $Q^{t+1}_i$ and $Q^{t+1}_j$
\\15 \hspace{0.75cm} {\bf end}
\\16 \hspace{0.75cm} $Q^{t+1}_i, Q^{t+1}_j$ $\longleftarrow$ $Sel^{t+1}_i, Sel^{t+1}_j$
\\17 \hspace{0.25cm} {\bf end}
\\18 \hspace{0.25cm} select best joint policy tree $\delta^T$ from $\{Q_i^t,Q_j^t\}$
\\19 \hspace{0.25cm} return $\delta^T$
\\{\bf end}
\\
\hline
\end{tabular}
\end{table}
One of drawbacks in the pruning process is that it cannot predict which beliefs about the state and about the other SUs' policies will eventually be useful before reaching the root of the policy trees. The MBDP algorithm combines the bottom-up and top-down approaches. By using top-down heuristics to
find out relevant belief states, the DP
algorithm compare the bottom-up policy trees and select the best joint policy. Some top-down heuristic policies are proposed in \cite{seuken}. It is obvious that the state of channel is not affected by actions of SUs. Therefore, we can easily compute the most probable beliefs using the initial belief and Markov models of channels.
The MBDP algorithm which is used in this paper is shown in Table \ref{table1}. 
The algorithm is written for two SUs $i$ and $j$. It can be rewritten to any arbitrary number of SUs. The parameter $maxTree$ denotes the number of policy trees that are used in exhaustive backup for constructing next stage. In other words, the size of set $Backup(Q_i^{t})$ is $|A||maxTree|^{|O|}$. To evaluate each pair $(q_i,q_j)$ with respect to the belief vector $b$, the concept of value vector in POMDP is employed\cite{Smallwood}. The expected sum of reward with respect to the belief $b$, $V^t(b)$, is computed by dot product of value vector and assumed belief:
\begin{equation}
V^t(b)=\displaystyle\sum_{s \in S}b(s)v^t(s)
\end{equation}
where $v^t$ is a $|S|$-dimensional vector. For a depth-$t+1$ joint policy trees $\delta^{t+1}=(q_1^{t+1},...,q_n^{t+1})$, the value vector is:
\begin{align}
v^{t+1}(s)&=R(s)\nonumber\\&
+\displaystyle\sum_{\vec{o}\in O} P(\vec{o}|s,\delta^{t+1})[\displaystyle\sum_{s'\in S}P(s',\vec{o}|s,\delta^{t+1})v^{t}(s',\delta^{t+1}(\vec{o}))]
\end{align}               
where $\delta^{t+1}(\vec{o})$ is the joint policy of subtrees selected by SUs
after observation vector $\vec{o}$.
In \cite{seuken}, it is proved that the MBDP algorithm has a linear time complexity
with respect to the horizon length.
\section{Problem Definition}
\subsection{System Model}
\begin{figure}[!t]
\centering
\includegraphics[width=3.5in]{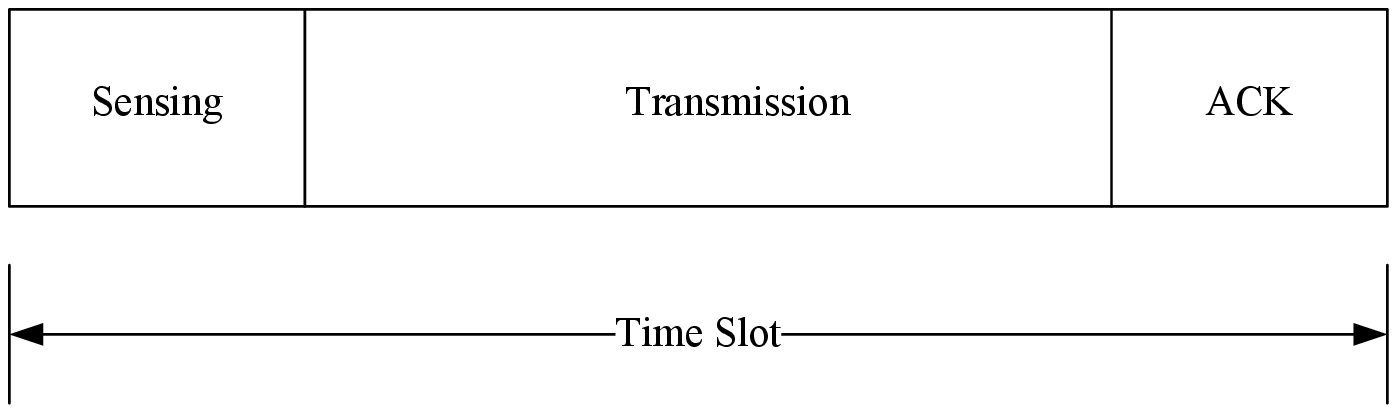}
\caption{A time slot}
\label{fign2}
\end{figure}

Our model consists of: $i$) A spectrum with $C$ channels, assigned to PUs; $ii$) $N$ PUs and $M$ SUs. It is assumed that all PUs and SUs communicate in a synchronous slot structure \cite{Zhao} and all SUs have same spectrum opportunities in each time slot (see Fig. \ref{fign1}). 
Each SU uses the beginning of each time slot to sense one of the $C$ channels. Based on the obtained observation, SUs can choose to either transmit on one
of the $C$ channels or not to transmit at all. At the end of the time slot, SUs receives ACK from their corresponding receiver to know that if transmission was successful or not (see Fig. \ref{fign2}).
\subsection{Dec-POMDP Formulation}
\begin{figure}[!b]
\centering
\includegraphics[width=3.1in]{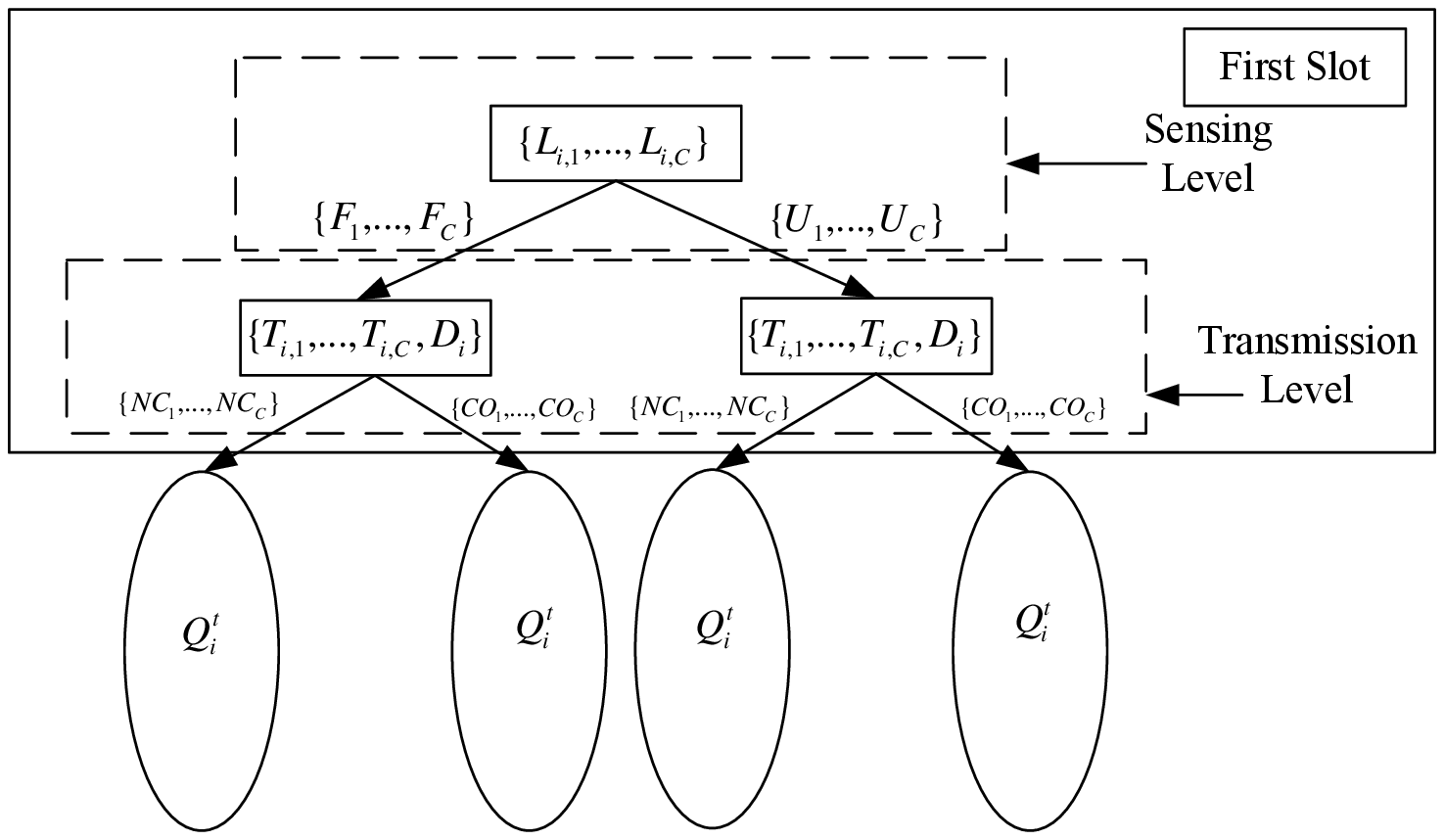}
\caption{Sets of depth-$t+1$ policy tree, $Q_i^{t+1}$ for $i$-th SU }
\label{fig2}
\end{figure}

For each SU, we have the following set of actions:  
\begin{equation}
A_i=\{L_{i,1},...,L_{i,C},T_{i,1},...,T_{i,C},D_i\}
\end{equation}
\\where $i=1,...,M$. $L_{i,j}$ represents the action of sensing channel $j$ for SU $i$. $L_{i,j}$s are only used in sensing level as shown in Fig. \ref{fig2}. $T_{i,j}$ denotes the action of accessing channel $j$ by SU $i$. It should be noted that $T_{i,j}$s are only used in transmission level as depicted in Fig. \ref{fig2}.
${D_i}$ shows the action that SU $i$ does not send its data and stays silent during the current time slot. Each channel is assumed to have a two state Markov model and also is independent of the other channels. Thus, we have:
\begin{equation}
S_i=\{0,1\}
\end{equation} 
\\where $S_i$ shows the set of states for channel $i$. Channel $i$ is available for SUs if $S_i=1$. The probability of state transition for Markov model of channel $i$ is represented by $p^i_{j,k}$ where $j,k\in\{0,1\}$.
For SU $i$, the following set of observations is considered:
\begin{equation}
O_i=\{U_i,F_i,CO_i,NC_i\}
\end{equation}
\\where $U_i$ shows that the channel $i$ is occupied by PUs and $F_i$ represents that channel $i$ is idle after sensing. In addition, $CO_i$ shows that there is a collision after transmission on channel $i$ and $NC_i$ denotes that there is no collision after action $T_i$. $CO_i$ and $NC_i$ are obtained through ACK signal sent by receiver. $U_i$ and $F_i$ are observations in sensing level while $CO_i$ and $NC_i$ are observations in transmission level.

We define $R_i(t)$, the reward function for SU $i$ in time slot $t$, as follows:
\begin{equation}
R_i(t)=\left\{ \begin{array}{ll}
  0  &\mbox{If SU $i$ does not send or recieves no ACK} \\
  1  &\mbox{If SU $i$ receives ACK}      
       \end{array} \right.
\end{equation}
It is clear that the reward function $R_i(t)$ depends on joint actions of SUs and states of channels. The reward function for Dec-POMDP formulation in time slot $t$ is obtained from the sum of all SUs' rewards:
\begin{equation}
R(t)=\displaystyle\sum_{i=1}^M R_i(t)
\end{equation}
 \begin{figure*}[t]
\centering
\includegraphics[width=5in]{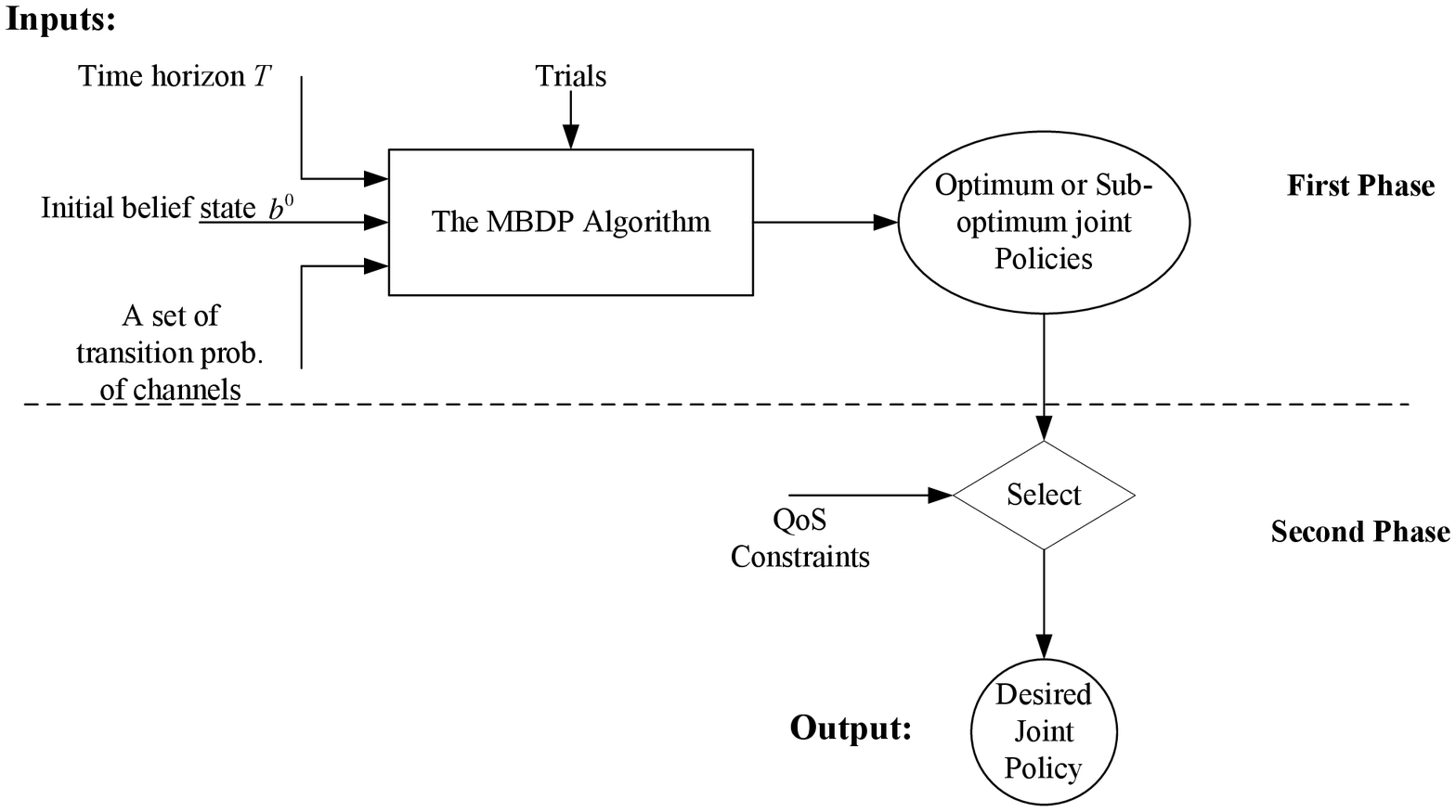}
\caption{The Proposed Method}
\label{fig3n}
\end{figure*}
\section{QoS-Aware Joint Policies}
In this section, we propose a method to find QoS-aware joint policies. The proposed method consists of two phases: 1- First Phase: In this phase, a set of optimum or sub-optimum joint policies are obtained using the MBDP algorithm. 2- Second Phase:  Among joint policies found in the first phase, the one which satisfies QoS constraints is selected.  
\subsection{Fisrt Phase}    
A set of optimum or sub-optimum joint policies are found in this phase by the MBDP algorithm. The MBDP algorithm has three inputs (see Fig. \ref{fig3n}): 1- time horizon ({\em T}): the duration of decision process. 2- initial belief state ($b^0$): the initial probability distribution of PUs' channel occupation. 3- A set of transition probabilities of channels: the transition probabilities of Markov chains ($p^i_{j,k}$s). The MBDP algorithm is executed for a specified number of trials.
\subsection{Second Phase}
In this phase, the joint policy that meets QoS constraints is selected. We define QoS constraints as the expected throughput of SUs which is given by the following equation:
\begin{equation}
\frac{R_1^T}{a_1}=\frac{R_2^T}{a_2}=...=\frac{R_M^T}{a_M}
\end{equation}
where $R_i^T$ is:
\begin{equation}
R_i^T=\mathbb{E}\left\{\displaystyle\sum_{t=1}^T R_i(t)\right\}
\end{equation}
and $\left\{a_i|i\in\{1,...,M\}\right\}$ is a set of QoS parameters.

{\em Definition}- We say joint policy $\delta^T$ satisfies QoS constriants if the $R_i^T$s meet following inequalities for an assumed parameter $\zeta$:
\begin{equation}
i=1,...,M,\exists t\in \left(0,(R_{max}+M\zeta)/\displaystyle\sum_{i=1}^Ma_i\right]:|R_i^T-a_it|\leq \zeta
\end{equation}
where $R_{max}$ is the maximum achievable throughput of cognitive radio network.
\subsection{Notes on Implementation Issues}
For different QoS constraints, optimum or sub-optimum joint policy trees are computed off-line and saved in the SUs' memory. Each joint policy tree has an identity number. Before SUs start sending, the initial belief is set to steady state distribution of channels. Moreover, SUs determine the initial belief precisely if they are allowed to sense all channels in the first slot. For the predefined horizon length ($T$) and assumed initial belief, SUs select an appropriate joint policy tree from their memory which guarantees QoS constraints. Afterwards, SUs send the identity number of the selected joint policy to their corresponding receivers in the first slot. The receiver tracks the transmitter's sensing action from decision tree and observes the channel which the transmitter is currently sensing. Because of the perfect sensing capability and same opportunistic spectrum for all SUs, the transceiver SUs are synchronized by knowing the selected joint policy tree.

As the horizon length is incremented by one, the number of leaves in decision tree increases exponentially. Therefore, saving trees requires large amount of memory. We assume that SUs transmit for a few number of slots in which saving decision trees is efficient. This assumption is reasonable in cases that statistical behavior of channels (e.g. the transition probability of Markov models) changes frequently and using a joint policy for very large horizon is not rational.
\begin{figure*}
    \centering
    \subfigure[SU1]
    {
        \includegraphics[width=7in]{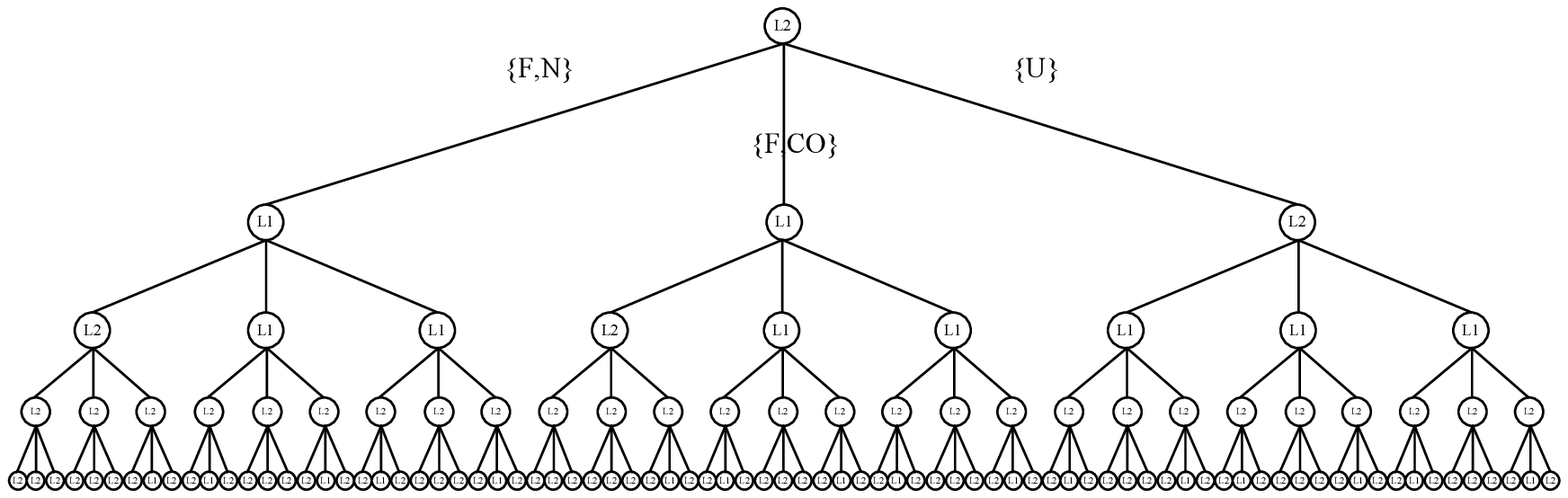}
        \label{fign5}
    }
    \subfigure[SU2]
    {
        \includegraphics[width=7in]{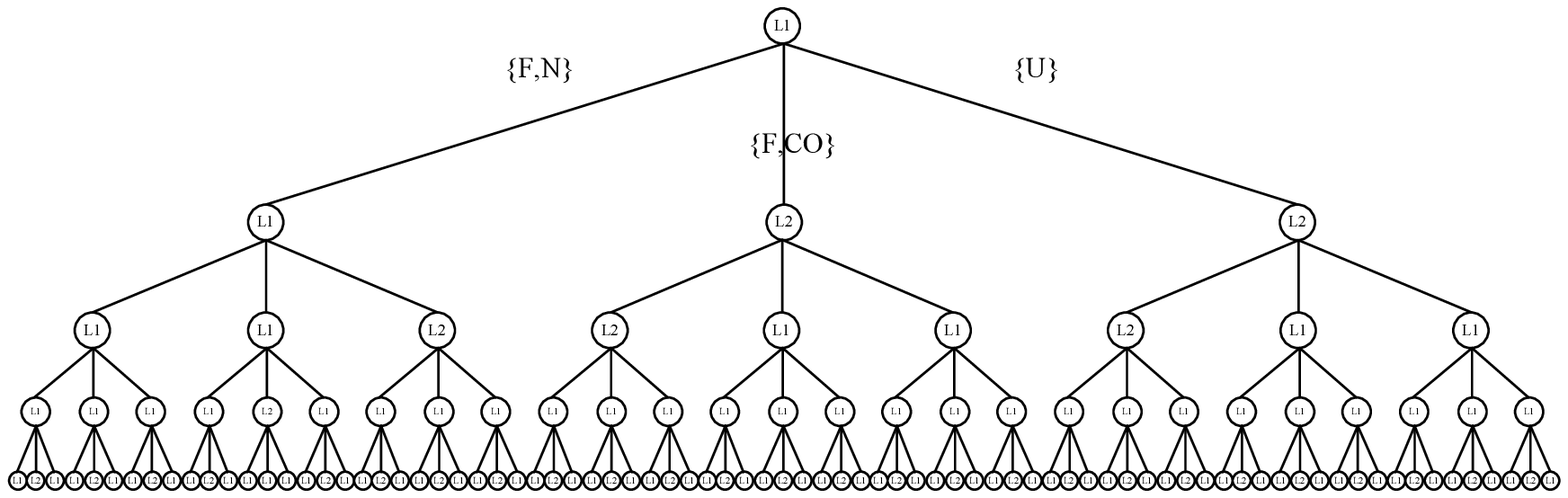}
        \label{fign6}
    }
	 \caption{Joint Policy tree}
\label{fig7}
\end{figure*}           
\section{Simulation}
To evaluate the performance of the proposed method, we consider two SUs in two-channel PUs' network. Each SU senses one of two channels in each time slot. The transition probabilities of channels are: $[p^1_{0,1},p^1_{1,0},p^2_{0,1},p^2_{1,0}]=[0.15,0.95,0.95,0.15]$. According to this distribution, channels $1$ and $2$ do not have same steady state probabilities. Channel $1$ is busy most of the time. It is also assumed that $S_1=1$ and $S_2=0$ in the first slot which determines the initial belief for the MBDP algorithm. Furthermore, each SU does not send and waits for next time slot if the observed channel is occupied by PUs. If the channel is idle, the SU sends. Besides, both SUs have perfect sensing capability. Considering above assumptions, we get a decision tree. An example of this tree is shown in Fig. \ref{fign5}. $L_i$ shows the action of sensing channel $i$. $\{F,N\}$ denotes that the observed channel is free and transmission is successful. $\{F,C\}$ denotes the observed channel is free but collision happens with other SU during transmission. $\{U\}$ means that observed channel is busy and the SU waits for the next time slot. An example of joint policy tree for $T = 5$ and $\frac{a1}{a2}=\frac{4}{3}$ is illustrated in Fig. \ref{fig7}. The set of observations for arcs in subtrees' root are as same as the original root. These observations are omitted because of simplicity of illustration.
   
The proposed method is compared with two related works on multi SU scheme\cite{Liu1,Liu2}. A multiuser heuristic (MH) policy for sensing channels is proposed in \cite{Liu1}. In other words, for SU $i$ with belief vector on availability of channels $\Omega^{(i)}(t)=[\omega_1^{(i)}(t),\omega_2^{(i)}(t)]$, the probability $p_n^{(i)}(t)$ of choosing channel $n$
in slot $t$ is given by:
\begin{equation}
p_n^{(i)}=\frac{\omega_n^{(i)}(t)}{\displaystyle\sum_{k=1}^2\omega_k^{(i)}(t)}
\end{equation}
In the other work \cite{Liu2}, it is assumed that SUs have different spectrum opportunities. A cooperation strategy which achieves near optimal throughput is given in \cite{Liu2}. In this strategy, SUs exchange their belief vector and use these information to take an action. If we rewrite cooperative approach formulation when SUs have same spectrum opportunities, the sensing strategy would be as follows:
\begin{equation}
{\small
\left\{ \begin{array}{ll}
SU1: L_1, SU2: L2&\mbox{If $\omega_1^{(1)}(t)+\omega_2^{(2)}(t)\geq\omega_1^{(2)}(t)+\omega_2^{(1)}(t)$} \\
SU1: L_2, SU2: L1&\mbox{Otherwise}      
       \end{array} \right.}
\end{equation}
    
\begin{figure}[!t]
\centering
\includegraphics[width=3.5in]{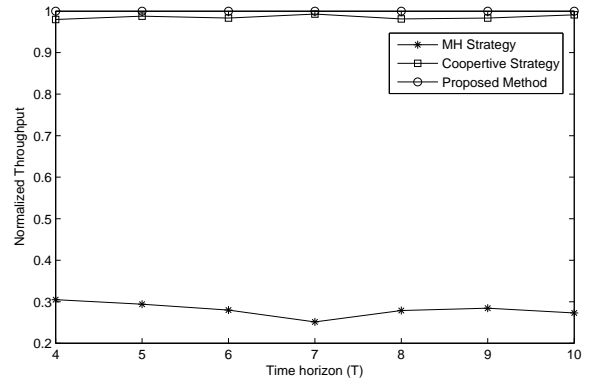}
\caption{Network Throughput}
\label{fign7}
\end{figure}

\begin{figure}
\centering
\includegraphics[width=3.5in]{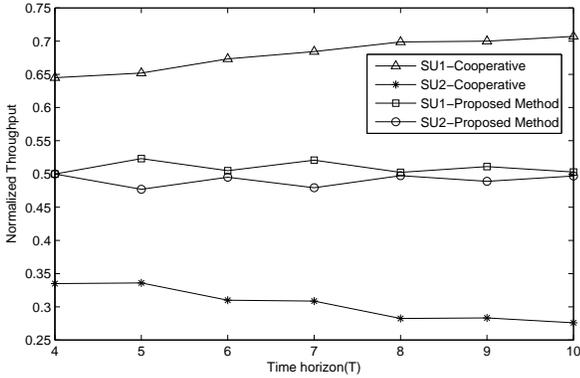}
\caption{SUs' Throughput (Fairness Comparison)}
\label{fign8}
\end{figure}

To find optimum or sub-optimum policies, we run the MBDP algorithm for 30 times in each horizon length ({\em T}). The parameter $maxTree$ is set to three. The results of network throughput is shown in Fig. \ref{fign7}. The results are normalized to maximum achievable throughput of network. This figure demonstrates that the proposed method outperforms \cite{Liu1,Liu2}. We also consider the case $a_1=a_2$. Parameter $\zeta$ is 0.25 in this simulation. The results for these settings are shown in Fig. \ref{fign8} for both proposed method and cooperative approach. In the cooperative approach, the second SU's throughput starves. However, SUs have approximately fair throughput in the proposed method. The expected throughput for two SUs and different QoS constraints are presented in Table \ref{table2}. Parameter $\zeta$ is set to 0.25. The pair in each entry denotes the throughput of SU$1$ and SU$2$, respectively. Numerical results show that the proposed method guarantees QoS constrains.               
\begin{table}[!t]
\renewcommand{\arraystretch}{1.3}
\caption{Achieved throughput for Simulated QoS Constraints}
\label{table2}
\centering
\begin{tabular}{|c|c|c|c|}
\hline
Horizon length(T) & $\frac{a_1}{a_2}=1$ & $\frac{a_1}{a_2}=1.5$ & $\frac{a_1}{a_2}=2$
\\
\hline
\hline
4 & 2,2 & 2.55,1.77 & 2.66,1.34
\\
\hline
5 & 2.6150,2.385	& 3,2 & 3.5,1.5
\\
\hline
6 & 3.03,2.97 & 3.7275,2.39 & 4,2
\\
\hline
7 & 3.645,3.355	&4.32,2.68	&4.56,2.44
\\
\hline
8&4.02,3.98&4.9,3.1&5.44,2.65
\\
\hline
9&4.6,4.4&5.5,3.5&5.86,3.14
\\
\hline
10&5.03,4.97&5.92,4.08&6.62,3.38
\\
\hline
\end{tabular}
\end{table}
\section{CONCLUSION}
We proposed a new method to guarantee QoS requirements for sensing-based protocols in multi SUs' network. By employing the MBDP algorithm, some optimum or sub-optimum joint policies are found and the one that satisfies QoS constraints is selected as the joint sensing strategy. Besides, the proposed method ensures transceiver SUs synchronization. For two SUs in a two-channel PUs' network, simulations demonstrated that the proposed method achieves maximum throughput. Moreover, results show that the proposed method guarantees different QoS constraints. 

\ifCLASSOPTIONcaptionsoff
  \newpage
\fi

\end{document}